OPEN

# Magnetization dynamics of weakly interacting sub-100 nm square artificial spin ices

Jose M. Porro[1,2,3*], Sophie A. Morley[4,5,6], Diego Alba Venero[1], Rair Macêdo[7], Mark C. Rosamond[8], Edmund H. Linfield[8], Robert L. Stamps[9,10], Christopher H. Marrows[4] & Sean Langridge[1]

**Artificial Spin Ice (ASI), consisting of a two dimensional array of nanoscale magnetic elements, provides a fascinating opportunity to observe the physics of out-of-equilibrium systems. Initial studies concentrated on the static, frozen state, whilst more recent studies have accessed the out-of-equilibrium dynamic, fluctuating state. This opens up exciting possibilities such as the observation of systems exploring their energy landscape through monopole quasiparticle creation, potentially leading to ASI magnetricity, and to directly observe unconventional phase transitions. In this work we have measured and analysed the magnetic relaxation of thermally active ASI systems by means of SQUID magnetometry. We have investigated the effect of the interaction strength on the magnetization dynamics at different temperatures in the range where the nanomagnets are thermally active. We have observed that they follow an Arrhenius-type Néel-Brown behaviour. An unexpected negative correlation of the average blocking temperature with the interaction strength is also observed, which is supported by Monte Carlo simulations. The magnetization relaxation measurements show faster relaxation for more strongly coupled nanoelements with similar dimensions. The analysis of the stretching exponents obtained from the measurements suggest 1-D chain-like magnetization dynamics. This indicates that the nature of the interactions between nanoelements lowers the dimensionality of the ASI from 2-D to 1-D. Finally, we present a way to quantify the effective interaction energy of a square ASI system, and compare it to the interaction energy computed with micromagnetic simulations.**

Artificial Spin Ice (ASI) systems are patterns of interacting ferromagnetic nanoelements whose particular geometry forces the ground state of the system to be magnetically frustrated, as not all the pairwise magnetic dipolar interactions between elements can be satisfied simultaneously[1–3]. These lithographically defined nanostructures mimic the behaviour of the spin-ice pyrochlores, where the arrangement of the rare earth magnetic moments leads them to lie in a frustrated state[4–6]. In turn, the pyrochlore crystals take the spin-ice name from water ice, due to the fact that the magnetic moments in the spin-ices map to the proton ordering in the molecules of water ice[7]. The basic ingredient of the ASIs is that the dimensions of the nanomagnets that form the array force them to have an Ising-like single-domain bistable behaviour of the magnetization, so that they can be treated as macrospins. These two-dimensional analogues of naturally-occurring spin-ice materials are being extensively studied, as their properties can be tuned at will by changing the nanomagnets' dimensions, materials and/or geometries[8], providing a huge landscape of possibilities to explore. One of the biggest advantages of ASIs with respect to the bulk spin-ices is the possibility to directly access experimentally, in real space, the microstates through a variety of techniques, such as magnetic force microscopy (MFM), photoemission electron microscopy (PEEM), and resonant transmission X-ray microscopy (TXM). Possible applications of ASIs range from their use in devices as return-point memories[9], magnetic cellular automata devices[10], or magnetic metamaterials[11], due to the possibility of creating and displacing magnetic monopoles (of Nambu type)[12] in the ASIs.

[1]ISIS Neutron and Muon Facility, Rutherford Appleton Laboratory, Chilton, OX11 0QX, United Kingdom. [2]BCMaterials, Basque Center for Materials, Applications & Nanostructures, 48940, Leioa, Spain. [3]Ikerbasque, Basque Foundation for Science, 48013, Bilbao, Spain. [4]School of Physics and Astronomy, University of Leeds, Leeds, LS2 9JT, United Kingdom. [5]Department of Physics, University of California, Santa Cruz, 95064 CA, USA. [6]Advanced Light Source, Lawrence Berkeley National Laboratory, 1 Cyclotron Road, Berkeley, CA, 94720, USA. [7]James Watt School of Engineering, Electronics and Nanoscale Engineering Division, University of Glasgow, Glasgow, G12 8QQ, United Kingdom. [8]School of Electronics and Electrical Engineering, University of Leeds, Leeds, LS2 9JT, United Kingdom. [9]School of Physics and Astronomy, University of Glasgow, Glasgow, G12 8QQ, United Kingdom. [10]Department of Physics and Astronomy, University of Manitoba, Winnipeg, Manitoba, R3T 2N2, Canada. *email: jm.porro@bcmaterials.net

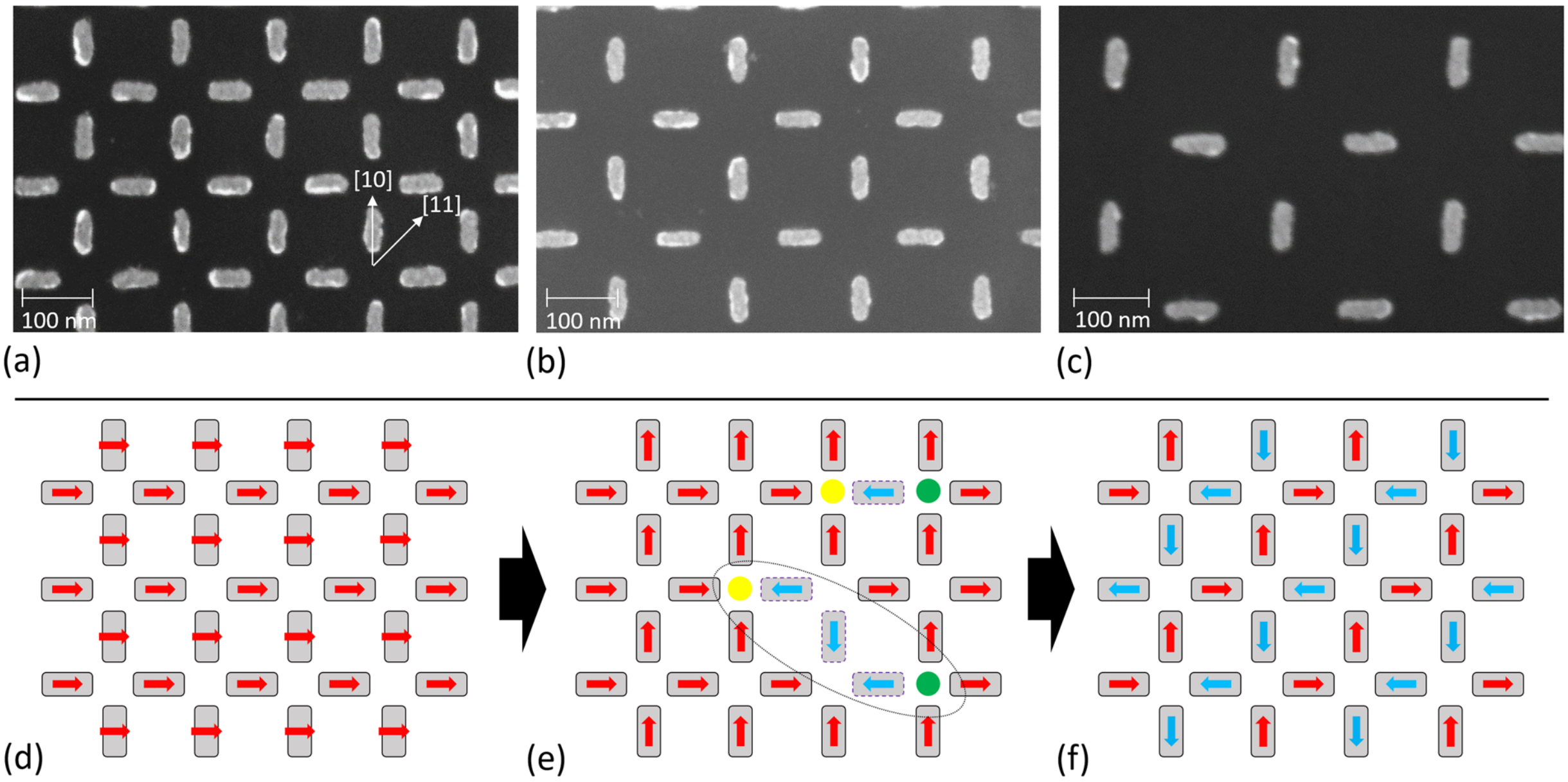

**Figure 1.** Square artificial spin ice and its magnetization dynamics process. Scanning electron microscopy images of square ASIs with three different lattice spacings: (**a**) 138 nm, (**b**) 175 nm and (**c**) 208 nm, made of Permalloy nanomagnets with lateral dimensions of 68 nm × 22 nm. Panel (a) shows the two directions [10] and [11] along which the field was applied whilst measurements were performed. The bottom panels (d–f) show schemes of the magnetization states in the pattern when a saturating field is applied right before starting the measurement (**d**); an intermediate state, at a certain time after starting the measurement, showing individual flips of nanomagnets that create monopolar charges (positive: yellow, and negative: green dots) connected by Dirac strings (encircled by a dashed ellipse in panel (e))[24] (**e**); and the final magnetization state, showing ground-state ordering of the square artificial spin-ice, after a full relaxation of the magnetization (**f**).

Until recently, studies on ASIs were performed on athermal systems, as the thermal energy needed to flip the magnetic macrospin of the nanomagnets forming the arrays was out of the experimentally accessible range. The studies on these athermal systems reported on effective thermodynamics, frozen excitations, and field demagnetization protocols in an attempt to access the ground-state ordering[13–19]. Recent reports on thermal ASIs have opened the door to the study of magnetization dynamics in these systems. These systems include thermal annealing processes taking place during fabrication[20], and systems where the anisotropy barrier of the nanomagnets has been tuned to be in a thermally accessible regime by judicious choice of a magnetic material with a lowered Curie temperature ($T_C$)[21,22] and by carefully heating the sample above its blocking temperature ($T_B$)[23]. These reports were shortly followed by studies of thermally fluctuating ASIs which have been imaged via PEEM[24–27] and TXM[28,29] in real time in a variety of geometries, and recently via MFM imaging of the intermediate thermally stable states after a temperature quenching process[30]. Nonetheless, it has only been very recently that experiments where a phase transition from the superparamagnetic regime ($T_C > T > T_B$) to the ASI regime have been demonstrated. These report glassy freezings of the magnetization dynamics of square ASI systems measured by X-ray photon correlation spectroscopy[31] and magnetometry[32], in both cases following a Vogel-Fulcher-Tammann law[33–35], a phenomenological law used to explain, among other systems, the behaviour of spin-glasses.

Building on the seminal MFM measurements of ASI, the huge advances in our knowledge of ASI have typically required access to intense x-ray synchrotron sources. In this context, we present a study to investigate the magnetization relaxation dynamics of square ASI systems by means of SQUID magnetometry. With this technique it is possible to explore the collective dynamics of the whole array of nanoelements composing the ASI, in comparison to previously mentioned techniques where only small portions of the sample are inspected with each measurement. The ASIs are formed by nanomagnets made of Permalloy (Ni80Fe20) with lateral sizes of 68 nm × 22 nm, with two different thicknesses: 5 nm and 6 nm; and three different lattice spacings for each thickness: 138 nm, 175 nm and 208 nm (Fig. 1), making a total of six samples studied. The symmetric disposition of the nanoelements avoid non-uniform magnetization states in them[36], ensuring their Ising-like behaviour. We observe that the relaxation dynamics of the studied square artificial spin-ices follow an Arrhenius-type Néel-Brown behaviour, contrary to what is reported previously in similar square artificial spin-ice studies[32]. Zero field cooling and field cooling measurements have been performed for all of the samples, together with magnetization relaxation measurements at fixed temperatures. The analysis of the data extracted from the measurements suggest 1-D magnetization dynamics processes. They also show a negative correlation of the average blocking temperatures of the samples with their interaction strength, indicating that the ASIs are in a weakly interacting regime, which is further corroborated following the theory of Shtrikman and Wohlfarth[37]. The experimental results are supported by Monte Carlo simulations of the magnetization processes in the samples studied. In systems of weakly interacting nanoparticles embedded in a polymeric matrix it has been observed that the blocking temperature of the system increases with the interparticle distances[38,39]; however, this is the first evidence for such a behaviour in a system where the interacting elements are not spherical nanoparticles, but elongated nanomagnets that are arranged in a pre-defined ordered geometry. In addition to the quantification of the interaction strength between the nanoelements in each ASI, the methodology followed here also gives information about the dimensionality of the system. Thus, the present work provides a systematic study of the effect of frustration in the dimensionality of artificial spin-ice systems with different geometries, and opens the door to the design and analysis of desired exotic states and emergent behaviours[40].

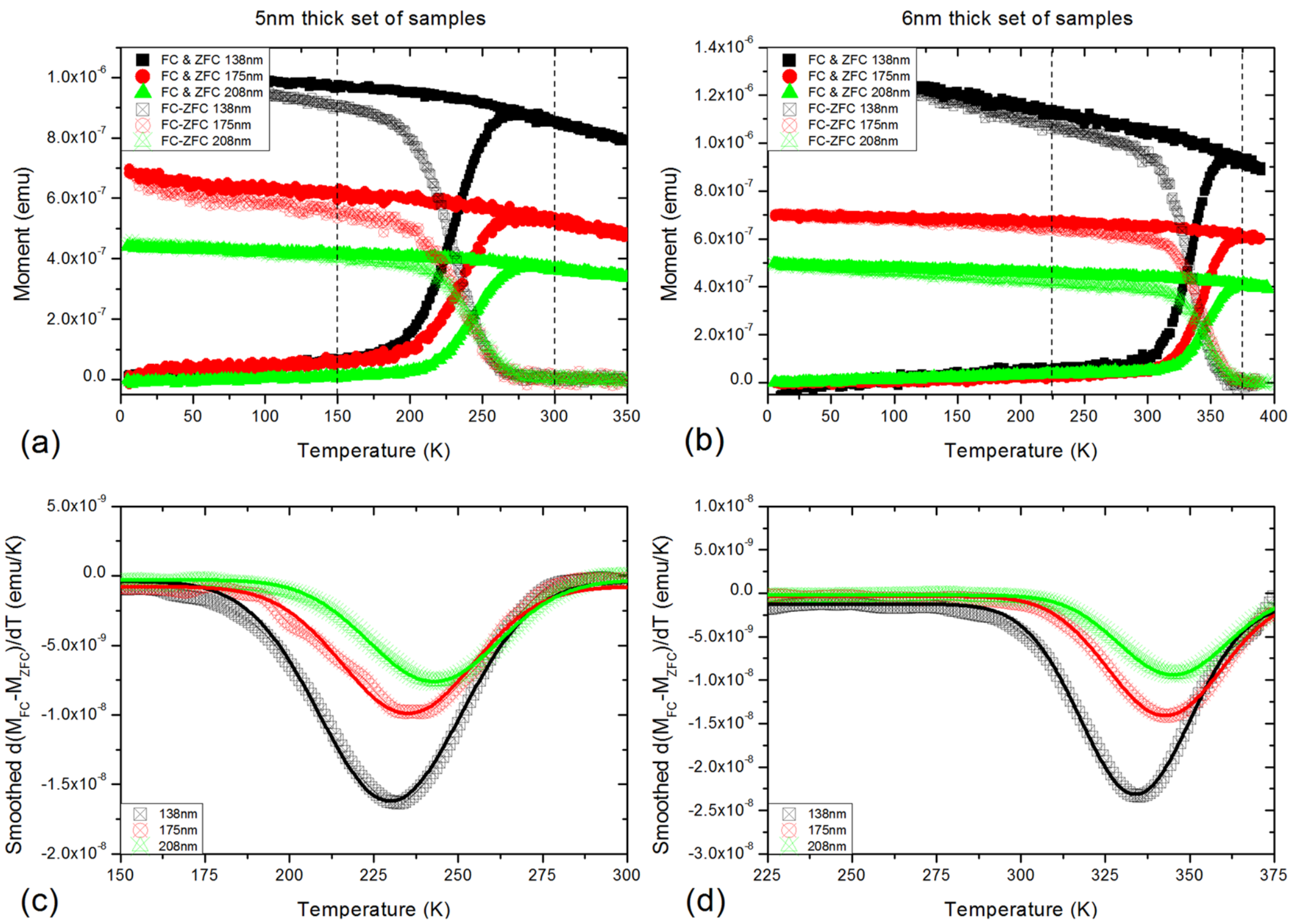

**Figure 2.** Zero field cooling/field cooling measurements. Solid symbols: zero field cooling/field cooling (ZFC/FC) measurements along the [10] directon on (**a**) the 5 nm thick samples set and (**b**) the 6 nm thick samples set, for each lattice spacing. Open symbols: subtraction of the zero field cooling curve from the field cooling curve (FC-ZFC) for each of the samples studied. (**c** and **d**) are the temperature derivatives of (FC-ZFC) for the 5 nm and 6 nm thick samples, respectively, where the blocking temperature $T_B$ in each sample corresponds to the position of the minimum in each curve. The line plots are smoothed curves from the datapoints. The vertical dashed lines in panels (a and b) denote the zoomed temperature regions plotted in (**c** and **d**), respectively.

## Results

**Zero field cooling/field cooling measurements.** In order to extract the characteristic relaxation times of the magnetization dynamics of our square ASIs we need to identify the temperature regions in which the samples are thermally active. This region of interest is readily identified using SQUID magnetometry. Details of the measurements can be found in the methods section. Zero field cooling (ZFC)/field cooling (FC) curves have been measured for all of the samples and are shown in panels (a) and (b) of Fig. 2. Upon cooling down in the absence of any external field, from a temperature above the average $T_B$ of the system, the nanomagnets will undergo slowing down of the magnetization dynamics until they reach a certain temperature below which the system will freeze into an ordered low energy ground state, where regions of nanomagnets arranged similarly to panel (f) of Fig. 1 are separated by higher energy vertex chains that form domain walls between ordered ground state regions[21]. The range of temperatures at which each sample will be thermally active is identified in the ZFC/FC measurements. The lower bound is given by the temperature at which the magnetization starts to increase in the ZFC (where we expect slow dynamics and long relaxation times), and the upper bound by the average $T_B$. It is expected that for temperatures slightly below the average $T_B$ fast dynamics and short relaxation times will be observed. From the lower branch (ZFC curves) of the plots in Fig. 2, the temperature range where the 5 nm thick samples are thermally active and with dynamics observable on laboratory timescales lies between 190 K and 270 K, as observed in panel (a), whereas for the 6 nm thick square ASIs it lies between 300 K and 380 K, as identified from panel (b). Note that above the upper bound temperatures specified here the samples are also thermally active, but their dynamics are so fast that we cannot experimentally probe them.

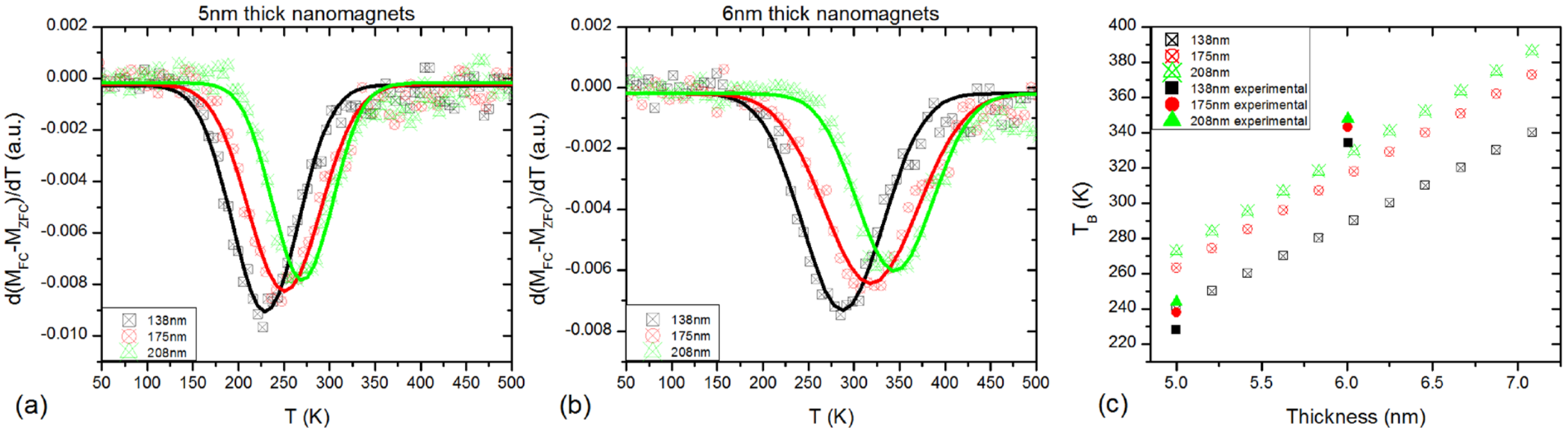

**Figure 3.** Monte Carlo simulations. Calculated temperature derivatives of the (FC-ZFC) curves for (**a**) the 5 nm and (**b**) 6 nm thick set of samples, using Monte Carlo simulation methods, for a direct comparison with insets (**c** and **d**) of Fig. 2. The solid lines are gaussian fittings to the data to determine the peak positions in each case. (**c**) Dependence of the $T_B$ of square ASI patterns formed by nanomagnets with the same lateral dimensions (68 nm × 22 nm) as a function of the thickness, for the three lattice spacings studied, calculated using Monte Carlo simulations. For comparison purposes, the average $T_B$ obtained from the experimental ZFC/FC measurements have also been included.

In order to observe the trend of the average $T_B$ for the samples studied we have plotted the difference between the FC and ZFC curves (FC-ZFC) for each measurement, shown as open symbols in panels (a) and (b) of Fig. 2. We have followed the methodology presented by Bruvera *et al.*[41] in order to determine the average blocking temperatures of our samples. For that purpose, we have plotted $d(M_{FC} - M_{ZFC})/dT$ for each sample, where the minimum of each curve corresponds to the average $T_B$ of the sample, and the results are shown in panels (c) and (d) of Fig. 2. From these plots, we can conclude that, while we would expect to observe a positive correlation between the interaction strength and the average $T_B$, our samples have a counter-intuitive behaviour of their average $T_B$: for both the 6 nm and 5 nm thick set of samples, the average $T_B$ is smaller for the smallest lattice spacing samples, whereas it is larger for the largest lattice spacing samples.

**Monte Carlo simulations.** To understand the remagnetization processes of square ASIs with similar dimensions and lattice spacings to the ones measured here we have performed numerical standard Metropolis Monte Carlo (MC) simulations (details in the methods section). We have simulated cooling processes from temperatures well above the average $T_B$ of each sample, under a small probe field of 30 Oe (similar to the field applied in the ZFC/FC measurements). For a direct comparison with the experimental results presented in panels (c) and (d) of Fig. 2, we have plotted, using the data obtained from the simulations, similar curves to those of that figure, which are presented in panels (a) and (b) of Fig. 3. The results obtained from these MC simulations match the experimental behaviour observed in our samples, as can be seen by comparing panels (c) and (d) of Fig. 2 with panels (a) and (b) of Fig. 3, respectively. For completeness, MC simulations have also been performed for samples with thicknesses between 5 nm and 6 nm, with the same lateral dimensions and lattice spacings as the ones studied in this manuscript. The average blocking temperatures obtained for each simulated square ASI are plotted in panel (c) of Fig. 3, as a function of the thickness. The average $T_B$ obtained from the experimental ZFC/FC measurements have also been included in panel (c) of Fig. 3 to allow for a direct comparison between experiments and simulations.

The results obtained for the MC simulations show an inverse correlation between the average $T_B$ of the samples and the strength of the dipolar interactions between the nanoelements, which is the same effect observed in the samples measured. Nonetheless, a constant temperature difference between the measured and simulated average $T_B$ exists for each set of samples: for the 5 nm thick set of samples the measured average $T_B$ are 22 ± 8 K lower than the calculated ones, while for the 6 nm thick set of samples they are 30 ± 10 K higher. These differences can be attributed to discrepancies between the nominal and real thicknesses of our samples, with the nominal 5 nm thick set of samples thinner than 5 nm, and the nominal 6 nm thick set of samples thicker than 6 nm. The differences between blocking temperatures of samples with different lattice spacings have been quantified, finding that there is a 2:1 ratio of $T_B$ differences between the 138 nm and 175 nm lattice spacing samples, and the 175 nm and 208 nm ones, both for the 5 nm and 6 nm thick set of samples. This 2:1 ratio of $T_B$ between different lattice spacings is also maintained in the simulations.

**Magnetization relaxation measurements.** After identifying the interesting region of temperature where each sample shows thermally activated dynamics we can measure the thermal relaxation of the magnetization dynamics in each sample for different fixed temperatures: for the 5 nm thick set of samples between 185 K and 265 K, in steps of 10 K (a total of 9 measurements for each sample); and for the 6 nm thick set of samples between 290 K and 370 K, also in steps of 10 K (again, 9 measurements for each sample). As an example, the recorded measurements of the average magnetization evolution in time, plotted in normalised form, at each fixed temperature for the 175 nm lattice spacing 6 nm thick sample are shown in Fig. 4. Each measurement is normalized to the magnetization saturation of each sample measured during the application of the saturating field before starting the magnetization relaxation measurement protocol. The normalized moment $m/m_S$ is fitted to a stretched exponential[42] of the form:

$$m(t)/m_S = \exp[-(t/t_r)^{\beta}], \quad (1)$$

where a characteristic relaxation time, $t_r$, of the magnetization dynamics and the stretching exponent, $\beta$, are extracted from each fit. The superimposed lines plotted on top of each measurement (scattered points plots) correspond to the fitted stretched exponential for that measurement. While the meaning of the extracted relaxation times for each sample at each temperature is of essential importance to identify the type of relaxation dynamics followed by the sample, the extracted stretching exponents also possess important information about the relaxation dynamics, as will be discussed.

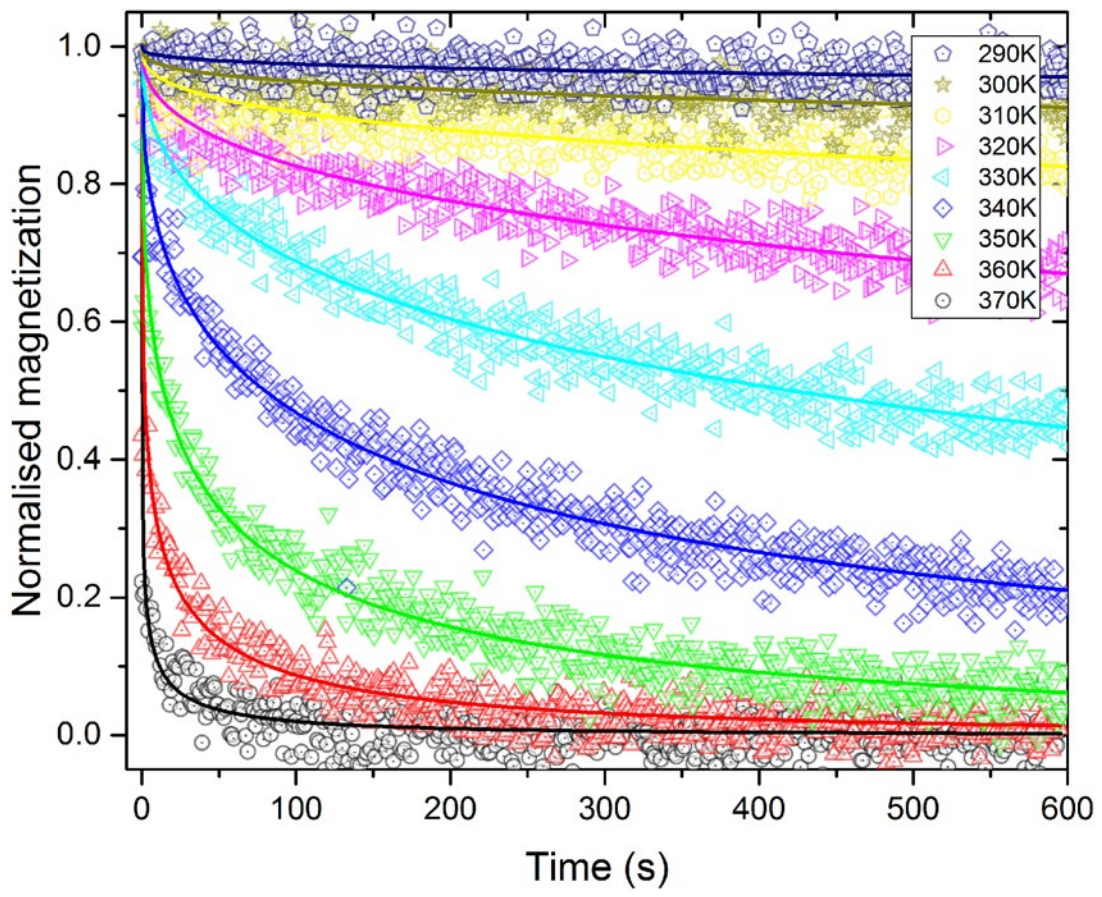

**Figure 4.** Magnetization relaxation experiments. Measurements of the magnetization relaxation dynamics of the 6 nm thick sample with 175 nm lattice spacing. Each scatter plot corresponds to the measurement of the time evolution of the magnetic moment of the sample at a fixed temperature. The moment is normalised to the magnetization saturation of the sample measured during the application of the saturating field before starting the relaxation measurement. The superimposed lines correspond to the stretched exponential model fitted to the data to extract the relaxation time, $t_r$, and the stretching exponent, $\beta$.

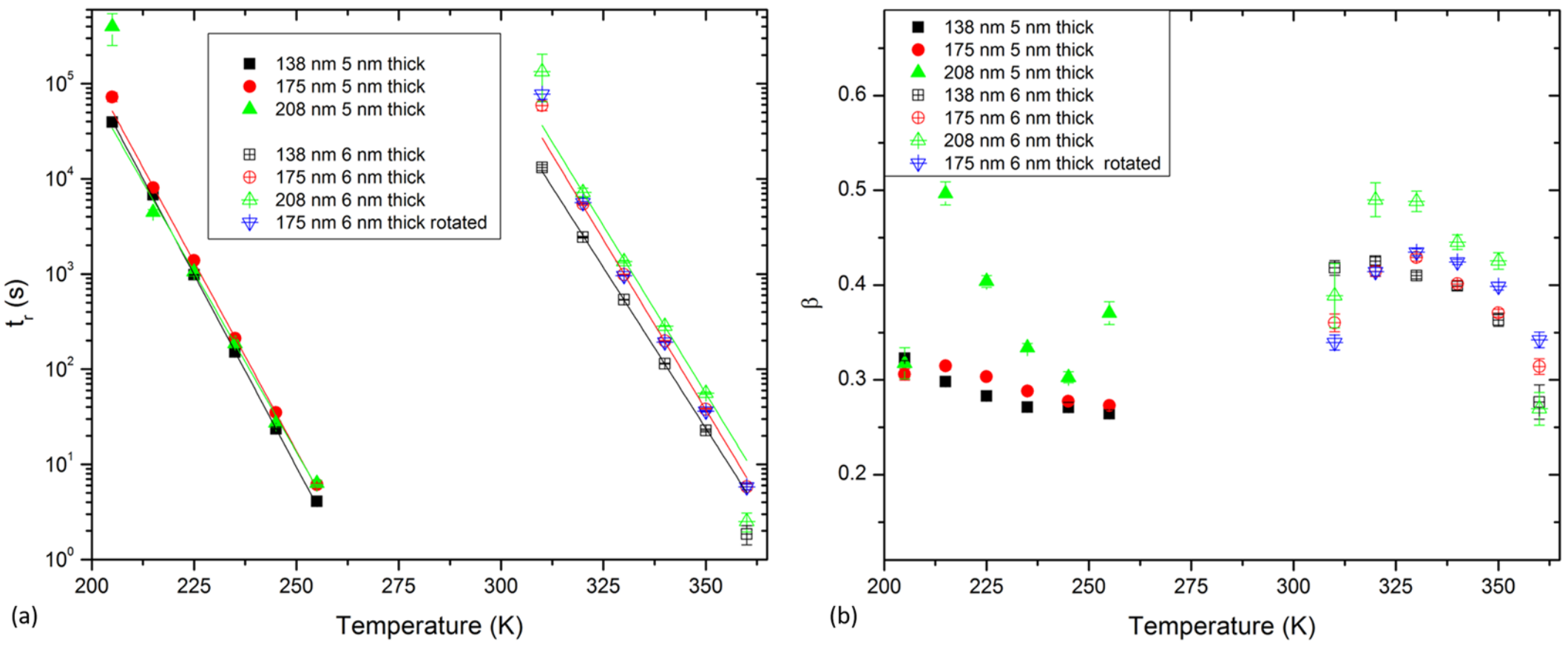

**Figure 5.** Relaxation times and stretching exponents of the samples studied. Temperature dependence of the relaxation times (**a**) and the stretching exponents (**b**) extracted from the stretched exponential fits of the measured time evolution of the magnetic moment on the 5 and 6 nm thick samples, for the three different lattice spacings. The measurements were performed along the [10] direction, except for the blue dataset, which corresponds to a measurement of the 6 nm thick 175 nm lattice spacing sample along the [11] direction.

The dependence of the relaxation times and stretching exponents on the temperature, as extracted from the fits of each measurement to Eq. 1 for the six samples studied here, are presented in panels (a) and (b) of Fig. 5, respectively. For the sake of comparison, similar relaxation measurements taken along the [11] direction of the 175 nm lattice spacing 6 nm thick sample (45 degrees from the easy axes of the nanoelements) are also included in Fig. 5.

| Lattice Spacing | $(E_A + E_{int})/k_B$ | Interaction Energy | Computed Dipolar Energy (OOMMF) |
|---|---|---|---|
| 138 nm | 9500 ± 300 K | $(9.6 \pm 0.3) \times 10^{-20}$ J | $2.1159 \times 10^{-20}$ J |
| 175 nm | 9900 ± 200 K | $(10.1 \pm 0.3) \times 10^{-20}$ J | $0.7770 \times 10^{-20}$ J |
| 208 nm | 9700 ± 400 K | $(9.9 \pm 0.6) \times 10^{-20}$ J | $0.3781 \times 10^{-20}$ J |

**Table 1.** Table showing the extracted $(E_A + E_{int})/k_B$ terms from fits to the Néel-Brown equation, from which the measured interaction energies for the 5 nm thick set of samples are obtained. The dipolar interaction energies calculated *via* micromagnetic simulations are also shown.

| Lattice Spacing | $(E_A + E_{int})/k_B$ | Interaction Energy | Computed Dipolar Energy (OOMMF) |
|---|---|---|---|
| 138 nm | 17100 ± 500 K | $(19.5 \pm 0.6) \times 10^{-20}$ J | $3.0442 \times 10^{-20}$ J |
| 175 nm | 18500 ± 400 K | $(21.4 \pm 0.6) \times 10^{-20}$ J | $1.1186 \times 10^{-20}$ J |
| 208 nm | 18500 ± 800 K | $(21.0 \pm 1.0) \times 10^{-20}$ J | $0.5430 \times 10^{-20}$ J |

**Table 2.** Table showing the extracted $(E_A + E_{int})/k_B$ terms from the fits to the Néel-Brown equation, from which the measured interaction energies for the 6 nm thick set of samples are obtained. The dipolar interaction energies calculated *via* micromagnetic simulations are also shown.

The temperature dependence of the relaxation times obtained for each of the samples are fitted to an Arrhenius-type Néel-Brown law:

$$t_r(T) = \tau_0 \exp(T_A/T), \tag{2}$$

where the activation temperature $T_A$, a measure of the energy barrier, depends on the anisotropy energy ($E_A$) of the island and the interaction energy ($E_{int}$) between elements. From this, the activation temperature is given by:

$$T_A = (E_A + E_{int})/k_B = (KV + E_{int})/k_B, \tag{3}$$

where $K = \mu_0 M_S^2 \Delta D/2$ is the anisotropy constant, and $\Delta D$ is the difference between the in-plane demagnetizing factors of the nanoelements[43].

For each sample, the fitting of the temperature dependence of the relaxation times to Eq. 2 gives us two parameters: the value for $(E_A + E_{int})/k_B$ and the attempt frequency of the nanoelements, $\tau_0$. For all the fits to the Néel-Brown equation, the obtained values for $\tau_0$ are of the order of $10^{-10}$–$10^{-12}$ s and are attributable to the flipping rate of the magnetic moments of the individual nanoelements at high temperature[44,45]. The extracted $(E_A + E_{int})/k_B$ values for the 5 nm thick set of samples are presented in Table 1 and the ones for the three 6 nm thick samples are presented in Table 2. From the fitted values, and calculating the shape anisotropy energy as previously mentioned, the value of the measured interaction energy is calculated for each of the samples. This interaction energy is compared to the extracted dipolar interaction energies computed via micromagnetic simulations[46], by subtracting the magnetostatic energies computed for an unfavourable alignment of the neighbouring $(NN + 2NN)$ macrospins from those obtained with a favourable alignment. The discrepancies between the computed and experimentally obtained interaction energies are due to discrepancies between the real volume of the nanomagnets in the samples and the nominal volume used in the micromagnetic calculations.

Even if the relaxation times of the three 5 nm thick samples have been fitted to the Néel-Brown law, a more detailed inspection of Fig. 5 suggests a different behaviour for the 208 nm lattice spacing sample than that of the 138 nm and 175 nm lattice spacing samples. The relaxation time is always shorter for the 138 nm lattice spacing sample than for the 175 nm one for each temperature measured; a trend that is observed, by looking at Fig. 5, not only in the samples with those lattice spacings of the 6 nm thick set of samples, but also in the 208 nm lattice spacing sample. Nevertheless, this trend is not followed by the 5 nm thick 208 nm lattice spacing sample with respect to the other two 5 nm thick samples. Due to the smaller magnetic signal produced by this sample, as it is the one with the least total magnetic material, the magnetic relaxation measurements are more noisy than with any of the other samples measured here. This led to higher correlated fitting parameters, causing the error bars to be relatively small, but the results are less trustworthy due to the higher noise observed in the measurements of this sample. The extracted value for the relaxation and exponent behaviour at $T = 205$ K is noteworthy: the sample is essentially static at this temperature, and consequently the poor fitting of the stretched exponential to that measurement yield non-realistic values for $t_r$ and $\beta$. This is reflected in the bigger error bars in $t_r$ and $\beta$ for the 205 K measurement of the 5 nm thick sample shown in Fig. 5.

## Discussion

### Zero field cooling/field cooling.

When studying superparamagnetic relaxation it is generally observed that $T_B$ increases as the average interaction strength between particles is increased[47,48]. In the ASIs studied here, where the nanomagnets have an Arrhenius-type Néel-Brown behaviour of the magnetization dynamics, the energy barrier between the two stable magnetization states of the nanoelements (which defines the individual $T_B$ of the nanoelement) is given by the sum of the shape anisotropy of the nanoelement (independent of the lattice spacing and common to the three samples on each set) and the interaction energy due to the dipolar magnetic interactions between neighbouring nanoelements. From the ZFC-FC measurements performed in the samples studied here, and supported by Monte Carlo simulations, we can clearly observe that their average $T_B$ decreases with the increase of the interaction strength between the nanoelements forming each sample, both for the 5 nm and 6 nm thick set of samples. This suggests that the samples are in a weakly interacting regime[38], as it is discused hereafter. The Monte Carlo simulations of the ZFC/FC protocols show that the average $T_B$ of the ASIs studied are very sensitive to small changes in the thickness of the nanoelements, as derived from the results exposed in panel (c) of Fig. 3.

In a study on systems of interacting ferromagnetic nanoparticles by Mørup and Tronc[38], they observe an inverse correlation between the average $T_B$ and the interaction strength, and they developed a model to explain this effect. The key ingredient of this model is that particles with uniaxial magnetic anisotropy are exposed to time-dependent dipolar fields coming from the neighbouring particles. At a certain time each particle is exposed to a dipolar field, while its magnetization fluctuates between the two stable states. Those fluctuations have frequencies of the order of $10^{10}$–$10^{12}\,s^{-1}$, while occasionally the magnetization vector will access the energy barrier for some value of the angle $\phi$ defined by its magnetic moment and the average dipolar field sensed by the particle. For some values of that angle, the energy barrier is lowered, leading to a decrease of the relaxation time due to the dipolar interactions. They derived an expression for the average $T_B$, which yields an inverse correlation of the average $T_B$ with the interaction strength. In our system, we have uniaxial particles (nanomagnets) that are subjected to the dipolar fields coming from the neighbouring nanomagnets. Furthermore, due to the nature of the lithography process in our system, there are very small random deviations of the alignments of the easy axes of the nanomagnets, leading to a similar effect to that accounted for in the model with $\phi$ and the dipolar fields.

**Magnetization relaxation.** According to previous studies of interacting superparamagnets[49], higher (lower) interaction energies are expected in samples with higher (lower) average $T_B$. Based on this, and with the behaviour of the average $T_B$ observed in our samples, it is expected that the effective interaction energies extracted from the magnetic relaxation measurements for the strongest interacting samples (smaller lattice spacings) will be smaller than the ones for the weaker interacting samples (bigger lattice spacings).

From the data presented in Table 1 it is observed that the extracted interaction energies for the three 5 nm thick samples overlap each other, as we are in a such a weakly interacting regime it is impossible to measure a distinct effect on this set of samples, based only on the interaction energies extracted from the relaxation measurements. This means that, although with static magnetometry measurements (ZFC/FC) we can observe differences in the behaviour of our samples, with magnetization dynamics measurements we cannot infer a different behaviour in these samples. Nonetheless, from the experimentally obtained values for the interaction energies of the 6 nm thick set of samples, presented in Table 2, it is clear that, while the 175 nm and 208 nm lattice spacing samples have similar overlapping interaction energies, the interaction energy of the 138 nm lattice spacing sample is smaller than the other two. This means that the difference in interaction strength between the 138 nm and 175 nm lattice spacing samples marks a threshold, below which the extracted interaction energies are indistinguishable between samples from the magnetization dynamics measurements. Furthermore, the fact that the interaction strength of the 138 nm lattice spacing sample is smaller than those of the 175 nm and 208 nm lattice spacing samples is in good agreement with the inverse correlation of the $T_B$ and their interaction strengths, obtained from the ZFC/FC measurements.

Following the theory of Shtrikman and Wohlfarth[37], we can establish a criterion to determine whether our samples are in a weakly interacting regime or not. This criterion is based on the comparison between the interaction energy of the system and the anisotropy energy of the single element, and states that if $E_{int} \gg E_A$ the samples are in a strongly interacting regime, whereas if $E_{int} \ll E_A$ the samples are in a weakly interacting regime. In our case, the interaction energies are around two orders of magnitude smaller than the anisotropy energy of the single islands composing the ASI arrays, thus meaning that we are in a weakly interacting regime in all the cases studied here. In previous studies of magnetization dynamics in square artificial spin ice systems[31,32], the lattice spacings and dimensions of the nanomagnets composing the samples studied there indicate that these samples should be in a weakly interacting regime according to the Shtrikman and Wohlfarth theory. However, they did not observe an inverse correlation of the average $T_B$ with the interaction strength. However, in those ASIs the activation volumes were much smaller than the nanoelement volumes, which is an effect previously reported in ASIs with larger nanoelements[30]. This means that the nanoelements are able to reverse at much lower temperatures than those expected for the full nanoelement volume barrier. The nanoelements of the study presented here have an activation volume much closer to the full nanoisland volume, so that the Shtrikman-Wohlfarth theory of being in a truly weakly interacting limit applies, hence we expect to observe an inverse correlation between $T_B$ and the interaction strength.

Comparing the experimentally obtained interaction energies with the calculated values of the dipolar energies for each sample, it can be seen that although they do not overlap, they are of the same order of magnitude and their discrepancies are attributed to a reduction in the volume of the real sample with respect to the values used for the micromagnetic computations.

It is worth noting that the calculations of the dipolar energies presented in Tables 1 and 2 have been obtained relaxing the magnetization state from a situation in which all the nanomagnets were in a perfect single-domain state, resulting in a canting of the magnetic moments along the edges of the nanoelements, but not involving any dynamic process. Therefore, we do not expect to observe an inverse correlation between the dipolar interaction energies extracted from the calculations and the interaction strengths, as this is a result of the magnetization dynamics processes in the samples, as discussed previously.

From an inspection of panel (a) of Fig. 5 it is evident that the measurements with the samples mounted along the [10] and [11] directions are indistinguishable, as the relaxation times match perfectly, as expected, due to the fact that the underlying magnetization dynamics processes are identical in the two sublattices forming the square ASI arrays.

The meaning of the stretching exponents obtained from the magnetization relaxation measurements is related to the dimensionality of the dynamic processes taking place. As a result of the lithography process there is a distribution of the energy barriers between the two stable magnetization states in each of the nanoelements, giving rise to a random distribution of energy potentials in the square array. This maps on to the so-called trapping model[50], allowing us to extract information about the dynamic processes from it. This model predicts that the stretching exponent obtained from the fits of the dynamics measurements to Eq. 1 takes the form:

$$\beta = \frac{d}{d+2}, \tag{4}$$

where $d$ is the dimensionality of the system.

The stretching exponents for the 5 nm thick set of samples (panel (b) of Fig. 5) scatter around values of $\beta = 1/3$, excluding the irregular behaviour of the 5 nm thick 208 nm lattice spacing sample. This value of $\beta = 1/3$ suggests 1-D dynamic processes, similar to those observed in square artificial spin-ice systems studied by PEEM[24,26]. These 1-D processes consist of the formation and propagation of chains of nearest neighbour nanomagnets undergoing reversal processes, being in the so-called string regime. The string and domain regime formation in thermally active ASIs was first shown by Farhan *et al.*[24] for strongly interacting arrays. More recent work comprised the study of both strongly interacting square ASIs, showing similar results, and weakly interacting ones where vertical string propagations occur[30]. Both cases present 1-D thermal excitations which happen in the initial part of the relaxation process or, compared to the present work, at lower temperatures. Based on this, vertical 1-D excitations are expected for the 5 nm thick samples, whereas 1-D diagonal strings are expected for the 6 nm thick samples, leading in both cases to a convergence of $\beta = 1/3$.

The more complex temperature dependence for the 6 nm thick set of samples is attributed to a mixed fluctuating state of nanoelements of both string formation and propagation processes, together with domain formations. This leads to a departure from the one dimensional behaviour, increasing the value of $\beta$ towards 1/2, which corresponds to a 2-D fluctuating system. The temperature dependence of the stretching exponents shows an average decrease in $\beta$ for higher temperatures in all the samples. This is attributed to the confinement of the fluctuations that become domain walls, as shown in a previous study by Budrikis *et al.*[51], leading to a recovery of the 1-D nature of the thermal excitations at higher temperatures and, therefore, to a convergence towards $\beta = 1/3$ of the stretched exponents. It is worth noting that the measurements at both edges of the temperature region studied have relaxation times that are in the limits of the detection of the technique, and the corresponding fits to the measurements have a higher $\chi^2$ value.

## Conclusions

To summarize, we have studied the magnetization dynamics of sub-100 nm square artificial spin ice samples with different thicknesses and lattice spacings by means of SQUID magnetometry. From the measurements we can conclude that the magnetization relaxation times obtained as a function of the temperature follow a simple Arrhenius-type Néel-Brown behaviour. This is expected from interacting superparamagnetic nanoparticles[52] that do not freeze into a glassy state[32].

The average blocking temperatures have an inverse correlation with the interaction strength of samples formed with nanomagnets with similar dimensions. These results are supported by Monte Carlo simulations. This unexpected negative correlation of the interaction strength with the average $T_B$ is a first-time observation in systems of elongated nanomagnets disposed in a non-random ordered geometry, although it has been observed previously[38,39] for interacting spherical randomly distributed ferromagnetic nanoparticles. In a study carried out by Mørup and Tronc[38] this effect is explained with a model that assumes uniaxial magnetic anisotropy in the weakly interacting nanoparticles that are exposed to dipolar fields from the neighbouring nanoparticles, both characteristics being found in the square ASIs studied here.

The magnetization relaxation measurements for each temperature are fitted to a stretched exponential function, from which we obtain a characteristic relaxation time and a stretching exponent. For each set of samples with the same thickness, the relaxation times have a positive correlation with the lattice spacing, the only exception being the least interacting sample (208 nm lattice spacing, 5 nm thick), whose nanoelements are in the limit of being non-interacting to super weakly interacting. The stretching exponent values give information about the dimensionality of the magnetization dynamics processes, and a value of $\beta = 1/3$ implies one-dimensional magnetization dynamics processes. Departures from $\beta = 1/3$ are attributed to mixed regimes of string formation and propagation processes with domain formation. The reduced dimensionality of the system, which shows 1-D magnetization processes in a 2-D ensemble of nanomagnets, is a direct consequence of the effects of the geometric frustration present in the square ASIs.

The experimental method proposed here is not only able to quantify the interaction energy of ASIs, but can also be applied in general to any ensemble of interacting nanomagnets. This is not restricted to ensembles following an Arrhenius-type Néel-Brown behaviour, but also to those following any other law (e.g. Vogel-Fulcher-Tammann) where an analytic expression for the temperature dependence of the relaxation times exist.

## Methods

### Growth and structural characterization.

The ASI samples studied here have been fabricated by means of electron-beam lithography, following standard procedures. Firstly, a layer of ZEP resist is spin-coated on a Si [100] substrate after cleaning the substrate. Then, standard exposure to the electron beam procedures are followed, to lithographically define the nanoelements with the desired lateral dimensions and lattice spacings, followed by a developing process of the resist after exposure by rinsing in a chemical developer. A thin layer of Permalloy (Ni80Fe20) is deposited onto the masked substrate, followed by a 2 nm thick Al cap (to prevent the samples from oxidation) and finally a lift-off process results in the square ASI patterns (Fig. 1). Surfaces of 2 mm × 2 mm were lithographically defined and covered by the patterns, to optimize the signal in the magnetometer.

**Magnetic characterization.** The magnetic characterization was performed using a commercially available Quantum Design SQUID magnetometer. Zero field cooling curves have been measured by heating the samples to 400 K (above TB) and bringing them to 10 K in a field less than 10 Oe (remanent field when not applying any field by the magnetometer), to ensure that the samples are in the lowest energy state (ground state depicted in panel (f) of Fig. 1), followed by a measurement of the moment on the sample, as a function of the temperature, from 10 K to 400 K, in a probe field of 30 Oe. Field cooling curves have been measured by heating the samples to 400 K, and then measuring their magnetic moment from 400 K to 10 K under a probe field of 30 Oe. The zero field cooling and field cooling curves presented in Fig. 2 have been measured with the fields applied and the samples mounted along the [10] direction, and with a heating/cooling rate of 2 K/min. The data shown in panels (c) and (d) of Fig. 2 have been smoothed following an adjacent averaging smoothing method, where the four neighbours of each data point are averaged, replacing the value of the data point by the new average value. Thermal relaxation measurements have been performed following this procedure: firstly, a saturating field of 5000 Oe was applied, forcing the magnetic moments to align with the applied field, resulting in a magnetic configuration of the array similar to the one depicted in panel (d) of Fig. 1. Then, the magnetic field is removed and the time evolution of the magnetization (in the absence of any external field) is measured for 600 s, resulting in graphs similar to the individual scattered plots shown in Fig. 4. Note that the magnetization plots shown in that figure are normalised to the magnetization saturation measured during the application of the saturating field, in order to perform the fitting of the stretched exponential function to extract the characteristic relaxation time. All the measurements performed in this study have been done with 1 s averaging time. The square patterns have been measured mounting the samples in such a way that the field was applied along the [10] (parallel to one of the sublattices of the square array) and the [11] (45 degrees from the [10]) directions, showing similar results (see blue and red datasets in Fig. 5).

**Micromagnetic simulations.** The micromagnetic simulations of the magnetostatic energies have been performed by means of OOMMF[46], assuming the nominal island sizes, with cell sizes of $2 \times 2 \times 1\,\mathrm{nm}^3$, well below the exchange length of Permalloy in every dimension, and considering nearest-neigbour and second nearest-neighbour interactions between elements. The material properties used are the ones defined for Permalloy in the OOMMF package by default. The use of cell sizes $2 \times 2 \times 1\,\mathrm{nm}^3$ is motivated by the fact that similar results are obtained with $1 \times 1 \times 1\,\mathrm{nm}^3$ cell sizes, and the former reduces drastically the computing time.

**Monte Carlo simulations.** The energetics and magnetisation processes of the system described in Fig. 1 were also theoretically investigated using a standard Metropolis Monte Carlo algorithm, assuming that the nanoelements forming the system are identical. Here, the spin Hamiltonian $\mathscr{H}$ has the form[53]:

$$\mathscr{H} = \mathscr{H}_{\mathrm{dip}} + \mathscr{H}_{\mathrm{app}} \tag{5}$$

denoting terms for the dipolar interaction and externally applied field, respectively.

We consider the magnetic nanoislands to be well-separated enough so that they can be considered as Ising-like spins and can be taken as point dipoles. In this case, the interaction between the magnetic moments is given by the expression[54]

$$\mathscr{H}_{\mathrm{dip}} = H_{\mathrm{D}} \sum_{i \neq j} s_i s_j \left[ \frac{\hat{\sigma}_i \cdot \hat{\sigma}_j}{|\vec{r}_{ij}|^3} - \frac{3}{|\vec{r}_{ij}|^5} (\hat{\sigma}_i \cdot \vec{r}_{ij})(\hat{\sigma}_j \cdot \vec{r}_{ij}) \right]. \tag{6}$$

The spin at the $i$-th site has a moment $S_i = M_s V s_i \hat{\sigma}_i$ where the unit vector $\hat{\sigma}_i$ represents the magnetisation direction, $M_s = 760$ A/m and $s_i = \pm 1$. Here, $H_{\mathrm{D}} = \mu_0 (M_s V)^2 / 4\pi l^3$ where $l$ is the lattice spacing and $V$ is the volume of the naoelements[55].

The effect of an external field, such as the one applied to the real system, can be calculated by evaluating

$$\mathscr{H}_{\mathrm{app}} = -BM_s V \sum_i s_i. \tag{7}$$

The method followed in order to perform the Monte Carlo simulations is similar to those employed in literature in the context of nanoparticles and fine magnetic nanostructures[56–59] and, more in particular, in Artificial Spin-Ices[51,60,61]. The method is described as follows: for the field cooling simulation, the initial state at a temperature T well above TC of the ensemble consists of all the magnetic moments saturated and pointing towards the easy axis of one of the sublattice (the [10] direction described in Fig. 1). The standard Metropolis Monte Carlo method then calculates the equilibrium states for each temperature, where $10^5$ steps are sufficient for convergence, and once equilibrium is reached the temperature is decreased to the next simulation temperature. The temperature step used for the simulations is 7 K. A field of 30 Oe is applied during the simulation of the cooling process along the [10] direction. For the zero field cooling protocol, the initial state is the one depicted in panel (f) of Fig. 1 at a temperature of 10 K, and the temperature is increased in steps of 7 K up to 400 K under a probe field of 30 Oe. The components of the magnetization that contribute to M are those along the [10] direction. The stopping criterion for the simulation is reached when there are no significant spin fluctuations within each Monte Carlo step, corresponding to spin fluctuations of less than 2% in the averaged magnetization. The sensitivity of the Monte

Carlo simulations to the size of the chosen temperature step suffers from precision issues only close to the phase transitions. We performed test simulations with different temperature steps in order to determine the optimum temperature step size. In order to obtain averages for the Hamiltonian given in Eq. 5, we use a single-spin flip Monte Carlo approach. A single Monte Carlo step (MCS) is $L_x \times L_y$ single-spin flips, where $L_x$ and $L_y$ are the $x$ and $y$ lengths of the array (in terms of $l$), respectively. We have considered array sizes of $10^4$ spins.

## Data availability

The datasets generated and analysed during the current study are available from the corresponding author on reasonable request.

## Acknowledgements

This work was supported by the EPSRC (grant numbers EP/L003090/1, EP/L00285X/1, and EP/L002922/1). J.M. Porro acknowledges funding from the H2020 Excellent Science-Marie Skłodowska-Curie Actions with the individual fellow No. 753025.

## Author contributions

Sample growth: S.A.M. and M.C.R.; measurements: J.M.P. and S.A.M.; micromagnetic simulations: J.M.P.; Monte Carlo simulations: R.M.; analysis and interpretation: J.M.P., S.A.M., D.A.V., R.L.S., C.H.M. and S.L.; supervision of the project: E.H.L., R.L.S., C.H.M. and S.L. All authors contributed to the manuscript.

## Competing interests

The authors declare no competing interests.

## Additional information

**Correspondence** and requests for materials should be addressed to J.M.P.